\documentclass{elsart}
\usepackage{graphicx}
\usepackage{amsmath,amssymb}
\numberwithin{equation}{section}

\newcommand{\bbox}{\boldsymbol}
\newcommand{\be}{\begin{equation}}
\newcommand{\ee}{\end{equation}}
\newcommand{\bea}{\begin{eqnarray}}
\newcommand{\eea}{\end{eqnarray}}

\journal{Nuclear Physics A}

\begin{document}
\begin{frontmatter}
\title{Direct photons: a nonequilibrium signal of the expanding quark-gluon plasma}
\author[pitt]{S.-Y. Wang}, \ead{sywang@phyast.pitt.edu}
\author[pitt]{D. Boyanovsky\corauthref{cor}}, \ead{boyan@pitt.edu}
\corauth[cor]{Corresponding author.} and
\author[sinica]{K.-W. Ng}\ead{nkw@phys.sinica.edu.tw}
\address[pitt]{Department of Physics and
Astronomy, University of Pittsburgh, Pittsburgh, PA 15260, USA}
\address[sinica]{Institute of Physics, Academia Sinica, Taipei 115,
Taiwan, ROC}

\begin{abstract}
Direct photon production from a longitudinally expanding
quark-gluon plasma (QGP) at Relativistic Heavy Ion Collider (RHIC)
and Large Hadron Collider (LHC) energies is studied with a
real-time kinetic description that is consistently incorporated
with hydrodynamics. Within Bjorken's hydrodynamical model, energy
nonconserving (anti)quark bremsstrahlung $q(\bar{q})\rightarrow
q(\bar{q})\gamma$ and quark-antiquark annihilation
$q\bar{q}\rightarrow \gamma$ are shown to be the dominant
nonequilibrium effects during the transient lifetime of the QGP.
For central collisions we find a significant excess of direct
photons in the range of transverse momentum $1-2 \lesssim p_T
\lesssim 5$ GeV/$c$ as compared to equilibrium results. The photon
rapidity distribution exhibits a central plateau. The transverse
momentum distribution at midrapidity falls off with a {\em power
law} $p^{-\nu}_T$ with $2.5 \lesssim \nu \lesssim 3$ as a
consequence of these energy nonconserving processes, providing a
distinct experimental {\em nonequilibrium signature}. The power
law exponent $\nu$ increases with the initial temperature of the
QGP and hence with the total multiplicity rapidity distribution
$dN_\pi/dy$.
\end{abstract}

\end{frontmatter}

\section{Introduction}

The first observation of direct photon production in ultrarelativistic heavy
ion collisions has been reported recently by the CERN WA98 collaboration in
${}^{208}$Pb$+{}^{208}$Pb collisions at $\sqrt{s}=158A$ GeV at the Super Proton
Synchrotron (SPS)~\cite{WA98}. Most interestingly, a clear excess of direct
photons above the background photons predicted from hadronic decays is observed
in the range of transverse momentum $p_T > 1.5$ GeV/$c$ in central collisions.
As compared to proton-induced results at similar incident energy, the
transverse momentum distribution of direct photons shows excess direct photon
production in central collisions beyond that expected from proton-induced
reactions. These findings indicate not only the experimental feasibility of
using direct photons as a signature of the long-sought quark-gluon plasma
(QGP)~\cite{mueller} but also a deeper conceptual understanding of direct
photon production in ultrarelativistic heavy ion collisions.

Unlike many other new phases of matter created in the laboratory, the formation
and evolution of the QGP in ultrarelativistic heavy ion collisions is
inherently a nonequilibrium phenomenon~\cite{geiger,FOPI}. Currently, it is
theoretically accepted that parton-parton scatterings thermalize quarks and
gluons on a time scale of about 1 fm/$c$ after which the plasma undergoes
hydrodynamic expansion and cools adiabatically down to the quark-hadron phase
transition. If the transition is first order, quarks, gluons, and hadrons
coexist in a mixed phase, which after hadronization evolves until freeze-out.
Estimates based on energy deposited in the central collision region at the BNL
Relativistic Heavy Ion Collider (RHIC) energies $\sqrt{s}\sim 200A$ GeV suggest
that the lifetime of the deconfined QGP phase is of order 10 fm/$c$ with an
overall freeze-out time of order 100 fm/$c$. Different types of signatures are
proposed for each different phase.

Of all the potential signatures of a QGP~\cite{signals}, direct photons and
dileptons emitted by the QGP, i.e., electromagnetic probes, are free of
hadronic final state interactions and can provide a clean signature of the
early stages of a thermalized plasma of quarks and gluons. Therefore a
substantial effort has been devoted to a theoretical assessment of the spectra
of direct photons and dileptons emitted from the
QGP~\cite{feinberg,shuryak,mclerran,kapusta,baier,ruuskanen,aurenche}.

The theoretical framework for studying direct photon production from a
thermalized QGP begins with an equilibrium calculation of the emission rate at
finite
temperature~\cite{mclerran,kapusta,baier,ruuskanen,aurenche,gale,lebellac}.
This rate is then combined with the hydrodynamic description to obtain the
total yield of direct photons produced during the evolution of the
QGP~\cite{mclerran,kapusta,ruuskanen,sollfrank,srivastava,alam}.

In a recent development, it was shown in Ref.~\cite{boyanphoton}
that direct photon production from a thermalized QGP with a finite
lifetime, when calculated using a real-time kinetic approach, is
strongly enhanced by energy nonconserving (anti)quark
bremsstrahlung $q(\bar{q})\rightarrow q(\bar{q})\gamma$ and
quark-antiquark annihilation $q\bar{q}\rightarrow \gamma$. This
nonequilibrium contribution has been missed by all the previous
equilibrium rate calculations in the
literature~\cite{mclerran,kapusta,baier,aurenche}. Most
importantly, this novel result poses a serious question on the
applicability of the equilibrium rate calculations to direct
photon production from a transient expanding QGP that is expected
to be created in ultrarelativistic heavy ion collisions at RHIC
and Large Hadron Collider (LHC) energies.

{\bf The goals of this article}. (i) We argue that the finite
lifetime of a transient QGP raises a conceptual inconsistency in
the calculation of direct photon production via an equilibrium
rate. (ii) We introduce a real-time kinetic description which
naturally accounts for the finite lifetime and nonequilibrium
aspects of the QGP. (iii) This real-time kinetic approach is
consistently combined with Bjorken's
hydrodynamics~\cite{bjorken,blaizot} to obtain the direct photon
yield for a QGP that is expected to be formed in central
collisions at RHIC and LHC energies.
(iv) We focus in particular on experimental signatures associated
with processes that would be forbidden by energy conservation in a
QGP of infinite lifetime.
(v) It is {\em not} our goal to assess photon production from the
hadronic phase {\em but} to compare the real-time kinetic
predictions for the QGP phase to those obtained from the
equilibrium calculations. Furthermore we focus on extracting
potential nonequilibrium signatures associated with the transient
and nonequilibrium aspects of the QGP phase.

Recently the {\em equilibrium} photon yield for a QGP with
hydrodynamic expansion, along with the photon production from the
hadronic phase has been used to estimate the total photon yield at
CERN SPS energies~\cite{alam}. This reference provides a thorough
comparison between the yields from the QGP obtained from the
equilibrium calculation, and that from the hadronic phase with the
result that they are of comparable order. Instead our goal is to
compare the nonequilibrium yield from the QGP to that obtained
from the equilibrium formulation and we show a substantial
enhancement from nonequilibrium effects indicating that the
nonequilibrium yield from the QGP will stand out over that from
the hadronic phase.

{\bf Brief summary of the main results}. We incorporate the
real-time kinetic approach with the hydrodynamical evolution of
the QGP to study direct photon production for central collisions
at RHIC ($\sqrt{s}\sim 200A$ GeV) and LHC ($\sqrt{s}\sim 5500A$
GeV) energies. We find that during the finite QGP lifetime of
order $\lesssim 10-30$ fm/$c$, energy nonconserving processes lead
to a substantial enhancement in the total photon yield for $1-2
\lesssim p_T \lesssim 5$ GeV/$c$ with a rapidity distribution that
is fairly flat for $|y| \lesssim \eta_{\rm cen}$, where
$|\eta|<\eta_{\rm cen}$ denotes the central rapidity region of the
QGP. These processes also lead to a total photon yield at
midrapidity that falls off with the transverse momentum as
$p^{-\nu}_T$ with $2.5 \lesssim \nu \lesssim 3$ in this region of
transverse momentum. Numerical studies reveal that this exponent
increases with the initial temperature of the thermalized QGP.
This power law is a result of the transient and nonequilibrium
aspects of the QGP and provides a distinct experimental
nonequilibrium signature.

This article is organized as follows. In Sec.~II we first review the usual
$S$-matrix approach to direct photon production from a QGP. We then argue that
this approach has shortcomings and is conceptually and physically incompatible
with photon production from an expanding QGP with a {\em finite lifetime}. In
Sec.~III Bjorken's hydrodynamical model combined with the $S$-matrix rate
calculation of photon production from the QGP is briefly summarized and the
incompatibility of these two approaches is highlighted. In Sec.~IV we introduce
the real-time kinetic approach to photon production from a longitudinally
expanding QGP which is consistently combined with Bjorken's hydrodynamics. This
section contains our main results and we compare these to those obtained from
the usual approach. Our conclusions are presented in Sec.~V. In the Appendix we
show that the {\em nonequilibrium} photon production rate is related to the
Fourier transform of the nonequilibrium expectation value of the quark
electromagnetic current correlation function.

\section{$\bbox{S}$-matrix approach and its shortcomings}\label{sec:usualapproach}

Production of direct photons from a QGP in thermal equilibrium has been studied
extensively~\cite{feinberg,shuryak,mclerran,kapusta,baier,aurenche} because of
its relevance as a clean probe of the QGP. Since photons interact
electromagnetically with quarks their mean free path is much larger than the
size of the QGP, hence they are not thermalized and leave the medium carrying
direct information from the QGP. Because of the smallness of the
electromagnetic coupling, the photon production rate is calculated to lowest
order in the electromagnetic interaction. This rate is obtained from the
$S$-matrix calculation of the transition probability per unit time and is
related to the Fourier transform of the thermal expectation value of the quark
electromagnetic current correlation
function~\cite{feinberg,mclerran,gale,lebellac}, which in turn is determined by
the imaginary part of the photon self-energy at finite
temperature~\cite{mclerran,gale,lebellac}.

In order to highlight the shortcomings of this formulation and  to establish
contact with the real-time kinetic approach to photon production introduced in
Sec.~IV, we now review some important aspects of the calculation.

Let us write the total Hamiltonian in the form
\begin{equation}
H = H_0+H_{\rm int},\quad H_0 = H_{\rm QCD}+H_{\gamma}, \quad
H_{\rm int} = e\int d^3x\,J^{\mu} A_{\mu},\label{totalH}
\end{equation}
where $H_{\rm QCD}$ is the full QCD Hamiltonian, $H_{\gamma}$ is the free
photon Hamiltonian, and $H_{\rm int}$ is the interaction Hamiltonian between
quarks and photons with $J^{\mu}$ the quark electromagnetic current, $A^{\mu}$
the photon field, and $e$ the electromagnetic coupling constant.

Consider that at some initial time $t_i$ the state $|i\rangle$ is an eigenstate
of $H_0$ with no photons. The transition amplitude at time $t_f$ to a final
state $|f,\gamma_{\lambda}({\vec p})\rangle\equiv
|f\rangle\otimes|\gamma_{\lambda}({\vec p})\rangle$, again an eigenstate of
$H_0$ but with one photon of momentum ${\vec p}$ and polarization $\lambda$, is
up to an overall phase given by
\begin{equation}
 S(t_f,t_i) = \langle f,\gamma_{\lambda}({\vec p})|U(t_f,t_i)|i\rangle,\label{transamp}
\end{equation}
where $U(t_f,t_i)$ is the time evolution operator in the interaction
representation
\begin{eqnarray}
U(t_f,t_i) &=& {\rm T}\,\exp\left[-i\int_{t_i}^{t_f}H_{{\rm int},I}(t)dt\right]\nonumber\\
&\simeq& 1-ie\int_{t_i}^{t_f}dt\int d^3x J^{\mu}_{I}({\vec x},t)
A_{\mu,I}({\vec x},t)+\mathcal{O}(e^2),\label{umatx}
\end{eqnarray}
where the subscript $I$ stands for the interaction representation in terms of
$H_0$. In the above expression we have approximated $U(t_f,t_i)$ to first order
in $e$, since we are interested in obtaining the probability of photon
production to lowest order in the electromagnetic interaction. The usual
$S$-matrix element for the transition is obtained from the transition amplitude
$S(t_f,t_i)$ above in the limits $t_i \rightarrow -\infty$ and $t_f \rightarrow
\infty$
\begin{eqnarray}
S_{fi}&\equiv&S(+\infty,-\infty)\nonumber\\
&=&-\frac{ie}{\sqrt{2E}}\int d^3x \int^{+\infty}_{-\infty}dt\,e^{iP^\mu
x_\mu}\,\varepsilon^{\lambda}_\mu\,\langle f|J^\mu(x)|i\rangle,\label{S}
\end{eqnarray}
where $E=|{\vec p}|$ and $P^\mu=(E,{\vec p})$ are the energy and four-momentum
of the photon, respectively, and $\varepsilon^{\lambda}_\mu$ is its
polarization four-vector. Since the states $|i\rangle$ and $ |f\rangle$ are
eigenstates of the {\em full} QCD Hamiltonian $H_{\rm QCD}$, the above
$S$-matrix element is obtained {\em to lowest order} in the electromagnetic
interaction, but {\em to all orders} in the strong interaction. We note that
the $S$-matrix element in effect is the amplitude for the transition between
asymptotic states $|i;{\rm in}\rangle\rightarrow|f,\gamma_{\lambda}({\vec
p});{\rm out}\rangle$, i.e., $S_{fi}=\langle f,\gamma_{\lambda}({\vec p});{\rm
out}|i;{\rm in}\rangle$, where $|f,\gamma_{\lambda}({\vec p});{\rm
out}\rangle\equiv|f;{\rm out}\rangle\otimes|\gamma_{\lambda}(\vec p);{\rm
out}\rangle$. Here, $|\gamma_{\lambda}({\vec p});{\rm out}\rangle$ is the
asymptotic {\sl out} state with one photon of polarization $\lambda$ and
momentum ${\vec p}$, and $|i;{\rm in}\rangle$ ($|f;{\rm out}\rangle$) is the
asymptotic {\sl in} ({\sl out}) state of the quarks and gluons.

The rate of photon production per unit volume from a QGP in thermal equilibrium
at temperature $T$ is obtained by squaring the $S$-matrix element, summing over
the final states, and averaging over the initial states with the thermal weight
$e^{-\beta E_i}/Z(\beta)$, where $\beta=1/T$, $E_i$ is the eigenvalue of $H_0$
corresponding to the eigenstate $|i\rangle$, and $Z(\beta)=\sum_i e^{-\beta
E_i}$ is the partition function. Using the resolution of identity $1=\sum_f
|f\rangle\langle f|$, the sum of final states leads to the electromagnetic
current correlation function. Upon using the translational invariance of this
correlation function, the two space-time integrals lead to energy-momentum
conservation multiplied by the space-time volume $\Omega=V(t_f-t_i)$ from the
product of Dirac delta functions. The term $t_f-t_i\rightarrow \infty$ is the
usual interpretation of $2\pi\delta(0)$ in the square of the energy conserving
delta functions.

These steps lead to the following result for the photon production rate in
thermal equilibrium~\cite{mclerran,ruuskanen}
\begin{eqnarray}
\frac{dN}{d^4x}&=&\frac{1}{\Omega}\frac{1}{Z(\beta)}\frac{d^3p}{(2\pi)^3}
\sum_{i,f,\lambda}e^{-\beta E_i}|S_{fi}|^2\nonumber\\ &=& -e^2\,
g^{\mu\nu}\,W^<_{\mu\nu}\frac{d^3p}{2E(2\pi)^3},\label{A}
\end{eqnarray}
where $W^<_{\mu\nu}$ is the Fourier transform of the thermal expectation value
of the current correlation function defined by
\begin{equation}
W^<_{\mu\nu}=\int d^4x\,e^{iP\cdot x}\,\langle J_\mu(0)J_\nu(x)\rangle_\beta.
\end{equation}
In the expression above $\langle\cdots\rangle_\beta$ denotes the thermal
expectation value. To lowest order in $e^2$ but to all orders in the strong
interactions, $W^<_{\mu\nu}$ is related to the retarded photon self-energy
$\Pi^R_{\mu\nu}$ by~\cite{gale}
\begin{equation}
e^2\,W^<_{\mu\nu}=\frac{2}{e^{E/T}-1}\,{\rm Im}\Pi^R_{\mu\nu}. \label{genfor}
\end{equation}
Thus, one obtains the (Lorentz boost) invariant photon production rate
\begin{equation}
E\frac{dN}{d^3p\,d^4x}=-\frac{g^{\mu\nu}}{(2\pi)^3}\,{\rm
Im}\Pi^R_{\mu\nu}\frac{1}{e^{E/T}-1}. \label{invarate}
\end{equation}

Kapusta {\it et al}.~\cite{kapusta} and Baier {\it et al}.~\cite{baier}
calculated this photon production rate at one-loop order and showed that the
processes that contribute to the energetic ($E\gg T$) photon production rate
are gluon-to-photon Compton scattering off (anti)quark $q(\bar{q})g\rightarrow
q(\bar{q})\gamma$ and quark-antiquark annihilation to photon and gluon
$q\bar{q}\rightarrow g\gamma$. Aurenche {\it et al}.~\cite{aurenche} have
recently extended the result to two-loop order. They found that the two-loop
contributions to photon production rate arising from (anti)quark bremsstrahlung
$qq(g)\rightarrow qq(g)\gamma$ and quark-antiquark annihilation with scattering
$q\bar{q}q(g)\rightarrow q(g)\gamma$ are of the same order as those at one
loop, and completely dominate the photon production rate at high photon
energies ~\cite{aurenche}. For two light quark flavors ($u$ and $d$), the
invariant direct photon production rate to two-loop order from a QGP in thermal
equilibrium at temperature $T$ is given by~\cite{kapusta,baier,aurenche}
\begin{equation}
E\frac{dN}{d^3p\,d^4x}\bigg|_{\text{eq}}
=\frac{5}{9}\frac{\alpha\alpha_s}{2\pi^2}\,T^2\,e^{-E/T}
\bigg[\ln\frac{0.23 E}{\alpha_s T} +\frac{16(J_T-J_L)}{\pi^3}
\left(\ln 2+ \frac{E}{3T}\right)\bigg],\label{eqrate}
\end{equation}
where $J_T\approx 1.11$ and $J_L\approx -1.07$~\cite{steffen}. We
would like to emphasize that the  thermal photon production rate
(\ref{eqrate}) and in general the result (\ref{genfor}) has two
noteworthy features that are important to our discussions below:
(i) The thermal rate is a static, time-independent quantity as a
result of the equilibrium calculation. (ii) Emission of high
energy photons is exponentially suppressed by the Boltzmann factor
$e^{-E/T}$.

We have reproduced the steps leading to Eq.~(\ref{invarate}), which is the
expression for the photon production rate used in all calculations available in
the literature, to highlight several important steps in its derivation in order
to compare and contrast to the real-time analysis discussed below.

\begin{itemize}
\item[(i)]{
The initial states $|i\rangle$ are averaged with the thermal
probability distribution at the initial time $t_i$ for quarks and
gluons. In the usual calculation this initial time $t_i\rightarrow
-\infty$ and the initial state describes the photon vacuum and a
thermal ensemble of quarks and gluons.}

\item[(ii)]{
The transition amplitude is obtained via the time evolution operator
$U(t_f,t_i)$ evolved up to a time $t_f$ and the transition amplitude is
obtained by projecting onto a state $|f\rangle$ at time $t_f$, which in the
calculation is taken $t_f \rightarrow \infty$. The sum over the final states
leads to the electromagnetic current correlation function averaged over the
initial states with the Boltzmann probability distribution, i.e., the thermal
expectation value of the current correlation function.}
\item[(iii)]{
Taking $t_f\rightarrow \infty$ and $t_i \rightarrow -\infty$ and squaring the
transition amplitude leads to {\em energy conservation} and an overall factor
$t_f-t_i$. The rate (transition probability per unit time per unit volume) is
finally obtained by dividing by $V(t_f-t_i)$. The important point here is that
taking the limit of $t_f-t_i \rightarrow \infty$ results in {\em two} important
aspects: energy conservation and an overall factor of the time interval
$t_f-t_i$. The resulting rate is independent of the time interval and only
depends on the photon energy (and obviously the temperature).}

\item[(iv)]{{\bf Main assumptions in the usual computations}. In order to compare
our methods and results with those obtained within the usual framework
described above, it is important to highlight the main assumptions that
are implicit in {\em all} previous calculations of photon production
from a thermalized quark gluon plasma and that are explicitly displayed
by the derivation above. Firstly, the  initial state at $t_i$ (which in
the usual calculation is taken to $-\infty$) is taken to be a thermal
equilibrium ensemble of quarks and gluons but the {\em vacuum state}
for the physical transverse photons. Furthermore, the buildup of the
photon population is {\em neglected} under the assumption that the mean
free path of the photons is larger than the size of the plasma and the
photons escape without rescattering. This assumption thus neglects the
prompt photons produced during the pre-equilibrium stage. Indeed,
Srivastava and Geiger~\cite{srivageiger} have studied direct photons
from a pre-equilibrium stage via a parton cascade model that includes
pQCD parton cross sections and electromagnetic branching processes. The
usual computation of the prompt photon yield during the stage of a {\em
thermalized} QGP assumes that these photons have left the system and
the computation is therefore carried out to lowest order in $\alpha$
with an initial photon vacuum state. Obviously keeping the
pre-equilibrium photon population results in higher order corrections
in $\alpha$. In taking the final time $t_f$ to infinity in the
$S$-matrix element the assumption is that the thermalized state is {\em
stationary}, while in neglecting the buildup of the population the
assumption is that the photons leave the system without rescattering
and the photon population never builds up. These assumptions lead to
considering photon production only the lowest order in $\alpha$, since
the buildup of the photon population will necessarily imply higher
order corrections. Although these main assumptions are seldom spelled
out in detail, they underlie all previous calculations of the photon
production from a thermalized quark gluon plasma.}

\item[(v)]{The main reason that we delve on the specific steps of the usual
computation and on the detailed analysis of the main assumptions
is to emphasize that there is a conceptual limitation of this
approach when applied to an expanding QGP of {\em finite
lifetime}. The current theoretical understanding suggests that a
thermalized QGP results from a pre-equilibrium partonic stage on a
time scale of order 1 fm/$c$ after the collision, hence for
consistency one must choose $t_i \sim 1$ fm/$c$. Furthermore,
within the framework of hydrodynamic expansion, studied in detail
below, the QGP expands and cools during a time scale of about 10
fm/$c$. Hence for consistency to study photons produced by a quark
gluon plasma in local thermal equilibrium one must set $t_f \sim
10$ fm/$c$. Hydrodynamic evolution is an {\em initial value
problem}, indeed the state of the system is specified at an
initial (proper) time surface (to be local thermodynamic
equilibrium at a given initial temperature) and the hydrodynamic
equations are evolved in time to either the hadronization or
freeze-out surfaces. The calculation based on the S-matrix
approach takes the time interval to infinity, extracts a {\em
time-independent rate} and inputs this rate, assumed to be valid
for every cell in the comoving fluid, in the hydrodynamic
evolution during a finite lifetime. }
\end{itemize}

As stated in the introduction, however, the QGP produced in ultrarelativistic
heavy ion collisions is intrinsically a transient and nonequilibrium state. It
is therefore of phenomenological importance to study nonequilibrium effects on
direct photon production from an expanding QGP with a finite lifetime with the
goal of establishing potential experimental signatures.

The current understanding of the QGP formation, equilibration, and subsequent
evolution through the quark-hadron (and chiral) phase transitions is summarized
as follows. A pre-equilibrium stage dominated by parton-parton interactions and
strong colored fields which gives rise to quark and gluon production on time
scales $\lesssim 1$ fm/$c$~\cite{geiger}. The produced quarks and gluons
thermalize via elastic collisions on time scales $\sim 1$ fm/$c$. Hydrodynamics
is probably the most frequently used model to describe the evolution of the
next stage when quarks and gluons are in local thermal equilibrium (although
perhaps not in chemical equilibrium)~\cite{bjorken,blaizot}. The hydrodynamical
picture assumes local thermal equilibrium (LTE), a fluid form of the
energy-momentum tensor and the existence of an equation of state for the QGP.
The subsequent evolution of the QGP is uniquely determined by the
hydrodynamical equations, which are formulated as an {\em initial value
problem} with the initial conditions specified at the moment when the QGP
reaches local thermal equilibrium, i.e., at an initial time $t_i \sim 1$
fm/$c$. The (adiabatic) expansion and cooling of the QGP is then followed to
the transition temperature at which the equation of state is matched to that
describing the mixed and hadronic phases~\cite{ruuskanen,srivastava,alam}.

Our main observation is that the usual computations based on S-matrix
theory  extract a time independent rate after taking the infinite time
interval, which  is then used in a calculation of the photon yield
during a {\em finite time} hydrodynamic evolution.

While we do not question the general validity of the results
obtained via the $S$-matrix approach, we here focus on the
signature of processes available during the {\em finite} lifetime
of the QGP and that would be forbidden by energy conservation in
the infinite time limit.

\section{Bjorken's hydrodynamical model}

In order to highlight the conceptual limitation of the $S$-matrix
calculation for direct photon production for a transient QGP we
now review the essential features of the hydrodynamic
description~\cite{bjorken,blaizot}.  For computational simplicity
we work within Bjorken's hydrodynamical model of longitudinal
expansion of the QGP~\cite{bjorken}, which is briefly summarized
in this section. The main assumption in Bjorken's model is
longitudinal Lorentz boost invariance in the central rapidity
region of the QGP. This is motivated by the observation that the
particle spectra for the secondaries produced in $p+$N and N$+$N
collisions exhibit a central plateau in the rapidity space near
midrapidity. For a longitudinally expanding QGP, it is convenient
to introduce the proper time $\tau$ and space-time rapidity $\eta$
variables defined by
\begin{equation}
\tau=\sqrt{t^2-z^2},\quad\eta=\frac{1}{2}\ln\frac{t+z}{t-z},
\end{equation}
where $t$ and $z$, respectively, are the time and spatial coordinate along the
collision axis in the center of momentum (CM) frame. The transverse spatial
coordinates will be denoted as ${\vec x}_T$, hence the space-time integration
measure is given by
\begin{equation}
d^4x=\tau\,d\tau\,d\eta\,d^2 x_T.\label{measure}
\end{equation}
Invariance under (local) longitudinal Lorentz boost implies that thermodynamic
quantities are functions of $\tau$ only and do not depend on $\eta$.

In Bjorken's scenario~\cite{bjorken,blaizot} the QGP reaches local thermal
equilibrium at a temperature $T_i$ at a proper time of order  $\tau_i\sim
1\,{\rm fm}/c$ after the maximum overlap of the colliding nuclei. The initial
conditions for hydrodynamical equations are therefore specified on a
hypersurface of constant proper time $\tau_i$. The equation of state for the
locally thermalized QGP is taken to be that of the ultrarelativistic perfect
radiation fluid (corresponding to massless quarks and gluons). The longitudinal
expansion is described by the scaling ansatz $v^z=z/t$, where $v^z$ is the
collective fluid velocity of the hydrodynamical flow and describes free
streaming of the fluid. Hence the space-time rapidity equals to the fluid
rapidity. In terms of $\tau$ and $\eta$ the scaling ansatz implies that the
four-velocity of a given fluid cell in the CM frame is given by
$u^{\mu}=(\cosh\eta,0,0,\sinh\eta)$ with $u^{\mu}u_{\mu}=1$. The conservation
of total entropy leads to adiabatic expansion and cooling of the QGP according
to the cooling law~\cite{bjorken,blaizot}
\begin{equation}
T(\tau)=T_i\left(\frac{\tau_i}{\tau}\right)^{1/3}.\label{coolinglaw}
\end{equation}
Hence, the QGP phase ends at a proper time
$\tau_f=\tau_i(T_i/T_c)^3$, where $T_c\sim 160$ MeV is the
quark-hadron transition temperature. At RHIC energies the initial
thermalization temperature is estimated to be $T_i\sim 200-300$
MeV, which entails that the lifetime of the QGP phase is of order
$\lesssim 10$ fm/$c$.

\subsection*{Direct photon production in the $S$-matrix approach}

As discussed in detail above, the usual ($S$-matrix) calculation of
direct photon production from a hydrodynamically expanding QGP proceeds
as follows~\cite{kapusta,ruuskanen,sollfrank,srivastava,alam}.
\begin{itemize}
\item[(i)]{
First the rate of direct photon production is calculated within the $S$-matrix
framework described in the previous section, leading to Eq.~(\ref{eqrate}) for
the invariant rate. This expression for the rate describes the photon
production rate in the {\em local rest} (LR) frame of a fluid cell in which the
temperature is a function of the proper time of the fluid cell. The rate in the
CM frame is obtained by a local Lorentz boost $E \rightarrow P^{\mu}u_{\mu}$
and the replacement $T(t)\rightarrow T(\tau)$:
\begin{equation}
\frac{dN}{d^2p_T\,dy\,\tau\,d\tau\,d\eta\,d^2x_{T}}\bigg|_{\rm CM}=
E\frac{dN}{d^3p\,d^4x}\bigg|_{\rm LR} [P^{\mu}u_{\mu},T(\tau)],\label{CMrate}
\end{equation}
where $p_T$ and $y$ are the transverse momentum and rapidity of the photon,
respectively. }
\item[(ii)]{
The direct photon yield is now obtained by integrating the rate over the
space-time history of the QGP, from the initial hypersurface of constant proper
time $\tau_i$ to the final hypersurface of constant proper time $\tau_f$ at
which the phase transition occurs. This leads to the following form of the
total direct photon yield in the CM frame for central collisions:
\begin{equation}
\frac{dN}{d^2p_T\,dy}\bigg|_{\rm CM}=\pi R_A^2
\int_{\tau_i}^{\tau_f}d\tau\,\tau
\int_{-\eta_{\rm cen}}^{\eta_{\rm cen}}d\eta 
\,E\frac{dN}{d^3p\,d^4x}\bigg|_{\rm LR}
[P^{\mu}u_{\mu},T(\tau)],\label{yieldSmtx}
\end{equation}
where $R_A$ is the radius of the nuclei and $-\eta_{\rm cen} <
\eta < \eta_{\rm cen}$ denotes the central rapidity region in
which Bjorken's hydrodynamical description is valid. The fact that
the $S$-matrix calculation for the rate results in a
time-independent rate (a consequence of taking the infinite time
interval, and hence assuming a stationary source as discussed
above) determines that the only dependence of the rate in the LR
frame on the proper time is through the temperature which is
completely determined by the hydrodynamic expansion.}
\end{itemize}

It is at this stage that the conceptual incompatibility between the $S$-matrix
calculation of the photon production rate and its use in the evaluation of the
total photon yield from an expanding QGP of {\em finite lifetime} becomes
manifest.

The hydrodynamic evolution is treated as an initial value problem with
a distribution of quarks and gluons in {\em local thermal equilibrium}
on the initial hypersurface of constant proper time $\tau_i \sim 1$
fm/$c$. The subsequent evolution determines that the QGP is a transient
state with a lifetime of order $\lesssim 10$ fm/$c$. The direct photon
yield is obtained by integrating the rate over this {\em finite
lifetime}. The $S$-matrix calculation of the rate for a QGP in thermal
equilibrium, on the other hand, implicitly assumes that $\tau_i
\rightarrow -\infty$ and $\tau_f \rightarrow \infty$ as discussed above
in detail. Therefore {\em while the rate has been calculated by taking
the time interval to infinity assuming a stationary source, it is
integrated during a finite time interval to obtain the total yield}.

The question that we now address, which is the focus of this article, is the
following: {\em Is this conceptual incompatibility of physical relevance and if
so what are the experimental observables?}

In order to answer this question and to assess the potential experimental
signatures from nonequilibrium effects, we must depart from the $S$-matrix
formulation and provide a {\em real-time} calculation of the direct photon
production rate based on nonequilibrium quantum field theory.

\section{Real-time kinetic approach}

Recently there has been substantial progress in the real-time approach to
quantum kinetics in nonequilibrium quantum field
theory~\cite{boyanmarkov,boyankin,boyanrgkin,boyanQED}. The advantage of this
approach is that it allows to study the time evolution of the single
(quasi)particle distribution functions for the relevant degrees of freedom
directly in real time as an {\em initial value problem}. The initial conditions
for this initial value problem are specified in terms of the initial state of
the plasma at the onset of the evolution. Furthermore, in the weak coupling
limit a {\em perturbative} dynamical renormalization group method has been
developed to resum directly in real time the perturbative expansions for the
distribution functions. The corresponding real-time {\em dynamical
renormalization group equations} are the quantum kinetic (or quantum Boltzmann)
equations~\cite{boyanrgkin,boyanQED}.

This novel real-time kinetic approach has been applied to derive
quantum kinetic equations in hot scalar~\cite{boyanrgkin} and Abelian
gauge (scalar and spinor QED)~\cite{boyanrgkin,boyanQED} theories in a
gauge invariant manner.

Compared to the usual approach to quantum kinetics in which transition
probabilities are computed using Fermi's golden rule and energy conservation,
this real-time approach has the following noteworthy advantages: (i) It is
capable of capturing energy nonconserving effects arising from the finite
lifetime of the plasma, as completed collisions are not assumed {\it a priori}.
(ii) Because inverse time acts as an infrared cutoff, the real-time kinetic
approach reveals clearly the infrared (threshold) singularities that lead to
anomalous nonexponential relaxation, thus transcending the usual quasiparticle
approximation~\cite{boyankin,boyanrgkin,boyanQED}. (iii) Since both the
real-time kinetic approach and hydrodynamics are formulated as an {\em initial
value problem}, they can be incorporated consistently on the same footing. This
last point proves very important in the hydrodynamic description of photon
production.

\subsection{Nonexpanding QGP}

We begin our discussion with the calculation of the invariant photon production
rate from a nonexpanding QGP. This calculation is relevant because the result
is interpreted as the invariant photon production rate in the local rest frame
of a fluid cell. The corresponding rate in the CM frame is obtained simply by a
local Lorentz boost and the direct photon yield is obtained by integrating the
rate over the space-time history of the QGP as explained in the previous
section.

Because of the Abelian nature of the electromagnetic interaction, we will work
in a gauge invariant formulation in which physical observables (in the
electromagnetic sector) are manifestly gauge invariant and the physical photon
field is transverse (for details see Ref.~\cite{boyanQED}).

The real-time kinetic approach begins with the time evolution of an initially
prepared density matrix (see Refs.~\cite{boyankin,boyanrgkin,boyanQED} and
references therein). Consistent with the hydrodynamical initial value problem,
we consider that at the initial time $t_i$ quarks and gluons are thermalized
such that the initial state of the QGP is described by a thermal density matrix
at a given initial temperature $T_i$.

In order to compare our results with those obtained from the usual $S$-matrix
calculation, we will assume that photons are not present at the initial time,
i.e., the initial state is the vacuum for physical transverse photons and
neglect the build up of the photon population. We emphasize that this is {\em
not} an extra assumption in our treatment, but is one of the main assumptions
in {\em all} previous calculations, we simply adopt this assumption for
comparison. This assumption justifies a calculation of the yield to lowest
order in $\alpha$. While the usual approach assumes that the photons produced
during the {\em pre-equilibrium} stage~\cite{srivageiger} had left the plasma
without building up a population, this assumption can be relaxed by allowing an
initial photon distribution and including the photon occupation number in the
``gain'' and ``loss'' terms in the kinetic
equations~\cite{boyanmarkov,boyankin,boyanrgkin,boyanQED}. Since the photons
that were produced in the pre-equilibrium stage are a result of electromagnetic
processes, the initial population will necessarily lead to higher order
corrections in $\alpha$.

Thus consistently with all previous calculations and for comparison reasons we
will assume no initial photon population and no population build up and compute
the yield to lowest order in $\alpha$.

Therefore consistently with the S-matrix approach described above,  the
initial density matrix $\rho_i$ is taken to be of the form
\begin{equation}
\rho_i= \rho_{\rm QCD}\otimes |0_{\gamma}\rangle\langle 0_{\gamma}|,\quad
\rho_{\rm QCD}=e^{-H_{\rm QCD}/T_i}.\label{densmatx}
\end{equation}

We remark that the assumption that the initial density matrix is
that of a thermalized system (after the strong interactions
thermalize quarks and gluons on a time scale $\sim 1~\mbox{fm}/c$)
underlies the program that studies the {\em equilibrium}
properties of the QGP. This is our {\em only} assumption, i.e.,
that of a thermalized QGP at an initial time scale $t_i \approx 1~
\mbox{fm}/c$ and is consistent with the general assumptions behind
the equilibrium program. Of course this assumption is elevated to
that of LTE, again consistent with the hydrodynamical description
of an expanding QGP as an initial value problem.

At any time $t$ later, the density matrix is given by
\begin{equation}
\rho(t)= e^{-iH(t-t_i)}\,\rho_i\,e^{iH(t-t_i)},\label{timedepdens}
\end{equation}
with $H$ the total (time independent) Hamiltonian given by
Eq.~(\ref{totalH}). Since the initial density matrix is assumed to
describe thermal equilibrium for quarks and gluons and therefore
commutes with $H_0$ defined in Eq.~(\ref{totalH}), it is
straightforward to find
\begin{equation}
\rho(t) = e^{-iH_0t}\,U(t,t_i)\,\rho_i\,U^{-1}(t,t_i)\,e^{iH_0t}
\end{equation}
with $U(t,t_i)=e^{iH_0t}e^{-iH(t-t_i)}e^{-iH_0t_i}$ being the time evolution
operator in the interaction representation given by Eq.~(\ref{umatx}).

We note that the upper and lower time limits in the unitary time
evolution operator are a consequence of studying the time
evolution of an initial state determined at $t_i$ up to a {\em
finite} time $t$. This is the usual ingredient in the study of
transition matrix elements in time-dependent quantum mechanics. In
Refs.~\cite{boyanrgkin,boyanQED} we have previously provided a
gauge invariant treatment of the real-time perturbative expansion
for the Abelian sector which we use here to cast the real-time
study of direct photon production directly in terms of gauge
invariant observables (see also Ref.~\cite{boyanphoton}).

Taking
\begin{equation}\label{compa}
\rho_{\rm QCD}= \frac{1}{Z(T_i)}\sum_j e^{-E_j/T_i}|j\rangle\langle j|,
\end{equation}
where $|j\rangle$ is the eigenstate of the full QCD Hamiltonian $H_{\rm QCD}$
with eigenvalue $E_j$ and $Z(T_i)=\sum_j e^{-E_j/T_i}$ is the initial partition
function corresponding to $H_{\rm QCD}$, we establish direct contact between
the real-time kinetic approach and the $S$-matrix approach discussed in the
previous section. We haste to add, however, that unlike in the $S$-matrix
formulation  the initial and final times in the real-time kinetic approach will
{\em not} be taken to $\mp \infty$, respectively. Keeping a finite interval of
time as befits the description of an expanding QGP as a transient
nonequilibrium state, as will be shown below, allows energy nonconserving
processes that cannot be captured by the $S$-matrix calculation.

Since photons escape directly from the QGP without further interaction, it is
adequate to treat them as asymptotic particles. The number operator ${\hat{
N}}({\vec p})$ that counts the number of photons of momentum ${\vec p}$ per
unit phase space volume is defined by
\begin{equation}
\hat{N}({\vec p})=\sum^2_{\lambda=1}a^\dagger_\lambda({\vec p}) a_\lambda({\vec
p}),\label{numberop}
\end{equation}
where $a_\lambda({\vec p})$ [$a^\dagger_\lambda({\vec p})$] is the annihilation
(creation) operator that destroys (creates) a photon of momentum ${\vec p}$ and
transverse polarization $\lambda$. These annihilation and creation operators
can be written as usual in terms of the transverse component of the photon
field and its conjugate momentum~\cite{boyanphoton,boyanQED}.

The number of photons per unit phase space volume at time $t$ is given by
\begin{eqnarray}
(2\pi)^3\frac{dN(t)}{d^3p\,d^3x} &\equiv& \mbox{Tr}\left[\rho(t)\,\hat{N}({\vec
p})
\right] \nonumber\\
&=& \mbox{Tr}\left[\rho_i \,\hat{N}({\vec p},t)\right],\label{numberoft}
\end{eqnarray}
where $\hat{N}({\vec p},t)$ is the Heisenberg number operator.

The invariant photon production rate is given by $E\,dN(t)/d^3p\,d^4x$ and is
obtained by using the Heisenberg equations of motion for the Heisenberg
operator $ \hat{N}({\vec p},t)$ (for details, see
Refs.~\cite{boyanphoton,boyanQED}). A systematic framework to obtain the
equation of motion is that of nonequilibrium quantum field theory in which the
relevant correlation functions are obtained from a path integral along a
contour in complex time~\cite{CTP}. A forward branch refers to the evolution
forward in time and a backward branch refers to evolution backwards in time,
and correspond, respectively, to the time evolution operator that pre- and
post-multiplies the initial density matrix in Eq.~(\ref{timedepdens}). This
method has been used previously in the treatment of quantum kinetic equations,
and we refer the reader to Refs.~\cite{boyanphoton,boyanQED} for details. The
invariant photon production rate is given by~\cite{boyanphoton,boyanQED}
\begin{eqnarray}
E\frac{dN(t)}{d^3p\,d^4x}&=&\lim_{t'\rightarrow t}
\left(\frac{\partial}{\partial t'}-iE\right)
\sum^{N_f}_{f=1}\frac{e\,e_f}{2(2\pi)^3}
\int\frac{d^3q}{(2\pi)^3}
\big\langle\bar{\psi}_f^-(-{\vec k},t){\vec\gamma}
\cdot{\vec A}^+_T({\vec p},t')\nonumber\\
&&\times\,\psi_f^-({\vec q},t)\big\rangle +{\rm
c.c.},\label{a:rate1}
\end{eqnarray}
where ${\vec k}={\vec p}+{\vec q}$. In this equation, $N_f$ is the number of
quark flavors, $e_f$ is the quark charge in units of the electromagnetic
coupling constant $e$, ${\vec A}_T$ is the transverse component of the photon
field, $\psi_f$ is the (Abelian) gauge invariant quark fields,
$\langle\cdots\rangle$ denotes the {\em full} nonequilibrium expectation value,
and the superscripts ``$\pm$'' refer to fields defined in the forward ($+$) and
backward ($-$) time branches~\cite{CTP}.

As mentioned above we assume that there are no photons initially and
that those that are produced escape from the plasma without building up
their population. Therefore the QGP is effectively treated as the
vacuum for photons consistently with  one of the main assumptions in
{\em all} calculations performed in the literature.

Consequently, the nonequilibrium expectation values on the right-hand
side of Eq.~(\ref{a:rate1}) are computed perturbatively to order
$\alpha$ and in principle to all orders in $\alpha_s$ by using
real-time Feynman rules and propagators. The photon propagators are the
same as those in the vacuum.

We shall further assume the weak coupling limit $\alpha \ll \alpha_s \ll 1$.
Whereas the first limit is justified and is essential for  the interpretation
of electromagnetic signatures as clean probes of the QGP, the second limit can
only be justified for very high temperatures, and its validity in the regime of
interest can only be assumed so as to lead to a controlled perturbative
expansion.

\begin{figure}[t]
\begin{center}
\includegraphics[width=3.5in,keepaspectratio=true]{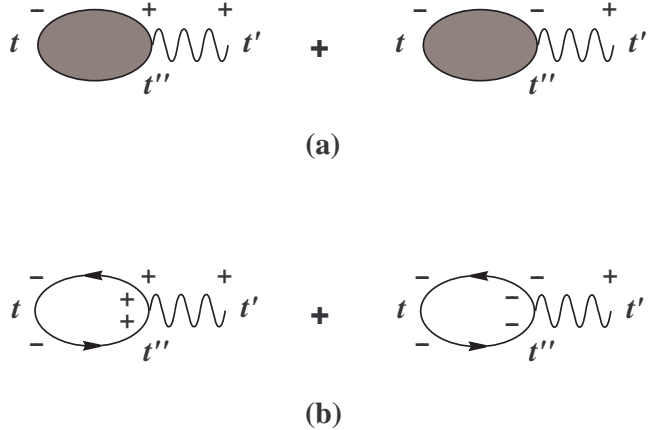}
\caption{Feynman diagrams for the invariant photon production rate
given by Eq.~(\ref{a:rate1}). Fig.~(a) is in terms of the full
self-energy of the photon to all orders in $\alpha_s$ and to order
$\alpha$. Fig.~(b) is the lowest order contribution, of order
$\alpha$.} \label{fig:loop}
\end{center}
\end{figure}

In the appendix we show that invariant photon production rate given by
Eq.~(\ref{a:rate1}) is related to the Fourier transform of the {\em
nonequilibrium expectation value} of the quark electromagnetic current
correlation function to lowest order in $\alpha$ and to {\em all orders} in
$\alpha_s$. The corresponding Feynman diagrams are shown in
Fig.~\ref{fig:loop}.

We focus on the {\em lowest order} (one-loop) contribution which, as has been
shown in Ref.~\cite{boyanphoton}, is missed by all the previous investigations
which use the $S$-matrix approach thus assuming an infinite QGP lifetime. In
using bare quark propagators we consider the quark momentum in the loop to be
hard, i.e., $q\gtrsim T_i$. Soft quark lines require hard thermal loop (HTL)
resummed effective quark propagator~\cite{brapis} leading to higher order
corrections. Indeed, the one-loop diagram with soft quark loop momentum is part
of the higher order contribution of order $\alpha\alpha_s$ that has been
calculated in Refs.~\cite{kapusta,baier}.

To {\em lowest order} in perturbation theory, the result for two
light quark flavors ($u$ and $d$) reads~\cite{boyanphoton}
\begin{eqnarray}
E\frac{dN(t)}{d^3p\,d^4x}&=&\frac{2}{(2\pi)^3} \int_{t_i}^{t}dt''
\int_{-\infty}^{+\infty}\frac{d\omega}{\pi}\,\mathcal{R}(\omega)
\,\cos[(\omega-E)(t-t'')],\label{a:noneqratei}
\end{eqnarray}
with
\begin{eqnarray}
\mathcal{R}(\omega)&=&\frac{20\,\pi^2\alpha}{3}\int\frac{d^3q}
{(2\pi)^{3}}\Big\{2\big[1-({\hat{p}}\cdot{\hat{k}})
({\hat{p}}\cdot{\hat{q}})\big] n(q)
[1-n(k)]\,\delta(\omega+k-q)\nonumber\\&&
+\big[1+({\hat{p}}\cdot{\hat{k}})
({\hat{p}}\cdot{\hat{q}})\big]
n(q)n(k)\,\delta(\omega-k-q)\Big\},\label{Ri}
\end{eqnarray}
where $q=|{\vec q}|$, $k=|{\vec p}+{\vec q|}$ and
$n(q)=1/[e^{q/T_i}+1]$ is the quark distribution function at the
{\em initial time} $t_i$. The dependence of $\mathcal{R}(\omega)$
on the quark distribution at the initial time $t_i$ is a
consequence of the fact that in nonequilibrium quantum field
theory the important ingredient is the {\em initial density
matrix}~\cite{CTP,boyankin,boyanrgkin,boyanQED}, which consistent
with hydrodynamics is taken to be thermal for quarks and gluons.
Obviously $\mathcal{R}(\omega)$ vanishes if there is no ``medium''
hence this contribution to photon production is a result of the
presence of the medium, which exists as a thermal bath for $t >
t_i \sim 1 ~\mbox{fm}/c$.

At this stage we comment on how the assumption of neglecting the photon
population can be relaxed. Allowing an initial as well as the buildup of the
photon population leads to a full quantum kinetic equation of the
form~\cite{boyankin,boyanrgkin,boyanQED}
\begin{equation}
\frac{d}{dt} n_{\gamma}({\vec p},t) = \Gamma_{\text{gain}}({\vec p},t)
[1+n_{\gamma}({\vec p},t)] -\Gamma_{\text{loss}}({\vec p},t)\,n_{\gamma}({\vec
p},t),
\end{equation}
where $n_{\gamma}({\vec p},t)= dN(t)/d^3 p d^3 x$,
$\Gamma_{\text{gain}}$ is given by the right hand side of
Eq.~(\ref{a:noneqratei}) divided by $E$,  and $\Gamma_{\text{loss}}$ is
obtained from $\Gamma_{\text{gain}}$ through the replacement
$n\leftrightarrow 1-n$ of the quark distribution functions (to the
order considered)~\cite{boyankin,boyanrgkin,boyanQED}. The population
at the time of onset of local thermodynamic equilibrium is a result of
electromagnetic processes during the pre-equilibrium
stage~\cite{srivageiger}, i.e., $n_{\gamma}({\vec p},t_0) \propto
\alpha$, thus leading to higher order corrections in $\alpha$.
Therefore, to lowest order in $\alpha$ the neglect of the $n_{\gamma}$
in the quantum kinetic equation is justified and leads to
Eq.~(\ref{a:noneqratei}).

In the case of a nonexpanding and thermalized QGP, the integral over $t''$ in
Eq.~(\ref{a:noneqratei}) can be carried out directly leading to
\begin{equation}
E\frac{dN(t)}{d^3p\,d^4x}=\frac{2}{(2\pi)^3}
\int_{-\infty}^{+\infty}d\omega\,\mathcal{R}(\omega)
\,\frac{\sin[(\omega-E)(t-t_i)]}{\pi(\omega-E)},\label{a:noneqratei2}
\end{equation}

A detailed analysis~\cite{boyanphoton,boyanQED} of
$\mathcal{R}(\omega)$ shows that the first delta function
$\delta(\omega+k-q)$ with support below the light cone
($\omega^2<E^2$) corresponds to the Landau damping cut, and the
second delta function $\delta(\omega-k-q)$ with support above the
light cone corresponds to the usual two-particle cut. Furthermore,
$\mathcal{R}(\omega)$ has a clear physical interpretation in terms
of the following {\em energy nonconserving} photon production
processes: the first term describes (anti)quark bremsstrahlung
$q(\bar{q})\rightarrow q(\bar{q})\gamma$, and the second term
describes quark-antiquark annihilation to photon
$q\bar{q}\rightarrow\gamma$. It has been shown in
Ref.~\cite{boyanphoton} that the dominant process for production
of energetic photons is the (anti)quark bremsstrahlung. As
explained in Ref.~\cite{boyanphoton}, in writing
$\mathcal{R}(\omega)$, we have ignored the term corresponding to
the ``vacuum'' process $0\rightarrow q\bar{q}\gamma$. The
``vacuum'' term, which persists for an infinitely long time,
results in an energy conserving delta function that obviously
vanishes.

At this stage we can make contact with the $S$-matrix calculation and highlight
the importance of the finite-time, nonequilibrium analysis. If, as is implicit
in the $S$-matrix calculation, the QGP is assumed in thermal equilibrium and
with an infinite lifetime (entailing that $t_i\rightarrow -\infty$ and $t_f \to
\infty$), then we can take the infinite time limit $t_i\to -\infty$ in the
argument of the sine function in Eq.~(\ref{a:noneqratei2}) and use the
approximation
\begin{equation}
\frac{\sin[(\omega-E)t]}{\pi(\omega-E)}\buildrel{t\rightarrow
\infty}\over{\approx}\delta(\omega-E).\label{goldenrule}
\end{equation}
This is the assumption of completed collisions that is invoked in
time-dependent perturbation theory leading to Fermi's golden rule
and energy conservation. The delta function $\delta(\omega-E)$ is
a manifestation of energy conservation for each {\em completed}
collision. Under this assumption one finds a {\em
time-independent} photon production rate proportional to
$\mathcal{R}(E)$, provided that the latter is finite. In the
present situation, however, the delta functions in
$\mathcal{R}(\omega)$ cannot be satisfied on the photon mass
shell. Therefore, under the assumption of completed collisions the
{\em lowest order} energy nonconserving contribution to the photon
production rate simply vanishes due to kinematics. Therefore this
lowest order contribution is absent (by energy conservation) in
the $S$-matrix calculation, but is present at any finite time.

The relevant question to ask is {\em how this finite-time contribution of order
$\alpha$ compares to the higher order $S$-matrix contribution to the photon
yield}.

The photon yield (per unit volume) is obtained by integrating the rate over the
lifetime of the QGP. Using Eq.~(\ref{a:noneqratei2}), one obtains
\begin{equation}
E\frac{dN}{d^3p~d^3x}=\frac{2}{(2\pi)^3} \int_{-\infty}^{+\infty}
d\omega\,\mathcal{R}(\omega)\,
\frac{1-\cos[(\omega-E)(t_f-t_i)]}{\pi(\omega-E)^2},\label{a:yield}
\end{equation}
where $t_f-t_i\sim10\,{\rm fm}/c$ is the lifetime of the QGP. For $T_i\sim 200$
MeV, the following important results have been shown in
Ref.~\cite{boyanphoton}.
\begin{itemize}
\item[(i)]{
Direct photon production given by Eq.~(\ref{a:yield}) features a power law
spectrum for $E\gg T_i$ as a consequence of the energy nonconserving photon
production processes: (anti)quark bremsstrahlung $q(\bar{q})\rightarrow
q(\bar{q})\gamma$ and quark-antiquark annihilation $q\bar{q}\rightarrow
\gamma$.}
\item[(ii)]{
The direct photon yield arising form this nonequilibrium finite-lifetime effect
in the energy range $E>2$ GeV dominates over that obtained from equilibrium
rate calculations given by Eq.~(\ref{eqrate}) during the QGP lifetime.}
\item[(iii)] {Furthermore the general analysis presented
in Refs.~\cite{boyanrgkin,boyanphoton} shows that the general
expression for the invariant rate given by Eq.~(\ref{a:yield}) in
terms of spectral representations is correct to  all orders. In
particular to lowest order in $\alpha$ but to all orders in
$\alpha_s$ the spectral density $\mathcal{R}(\omega)=
\alpha[\mathcal{R}^{(0)}(\omega)+ \alpha_s
\ln(1/\alpha_s)\mathcal{R}^{(1)}(\omega)+\cdots]$, all terms enter
with the {\em same time dependence}. In particular
$\mathcal{R}^{(0)}(\omega)$ is given by Eq.~(\ref{Ri}) and
$\mathcal{R}^{(1)}(\omega=E)$ is given by Eq.~(\ref{eqrate}). In
the limit $t_f-t_i \rightarrow \infty$ the terms with support at
$\omega =E$ result in a photon yield that grows {\em linearly with
time}, while the terms for which $\mathcal{R}(E)=0$ will grow in
time much slower. In the {\em infinite time limit} those terms
that do not vanish at $\omega =E$ lead to photon yield that grows
linearly with time and hence a finite time-independent rate, while
those that vanish will be subleading. However, for a finite time
interval which terms in the perturbative expansion dominate will
depend on the competition between the order in perturbation theory
and the length of the time interval. }
\end{itemize}

While these results in the nonexpanding case revealed the importance of
the nonequilibrium and finite-lifetime aspects, the most experimentally
relevant case to study is that of an expanding QGP. A constant rate
obtained from the $S$-matrix calculation leads to a yield that will
grow in time linearly, while as analyzed in Ref.~\cite{boyanphoton} the
lowest order nonequilibrium yield grows much slower (logarithmically).

Therefore the relevant question is  {\em which contribution is dominant
during the lifetime of the QGP }. This question can only be answered by
a detailed numerical study of both yields which is performed below.

\subsection{Longitudinally expanding QGP}

As a prelude to photon production from an expanding QGP, and according
with the usual approach~\cite{srivastava,alam}, we first focus on the
invariant photon production rate from each individual fluid cell of the
QGP. Since the proper time equals  the local time in the local rest
frame of any fluid cell, we can follow the same real-time
nonequilibrium analysis presented in the proceeding subsection to
calculate the nonequilibrium invariant photon production rate in the
local rest frame of each fluid cell~\cite{srivastava,alam}. We
obtain~\cite{boyanphoton,boyanQED}
\begin{eqnarray}
E\frac{dN(\tau)}{d^3p\,d^4x}\bigg|_{\rm
LR}&=&\lim_{\tau'\rightarrow\tau} \left(\frac{\partial}{\partial
\tau'}-iE\right) \sum^{N_f}_{f=1}\frac{e\,e_f}{2(2\pi)^3}
\int\frac{d^3q}{(2\pi)^3}
\big\langle\bar{\psi}_f^-(-{\vec k},\tau)\nonumber\\
&&\times\,{\vec\gamma}\cdot{\vec A}^+_T({\vec
p},\tau')\psi_f^-({\vec q},\tau)\big\rangle+{\rm
c.c.}.\label{rate1}
\end{eqnarray}

As before, the nonequilibrium expectation values on the right-hand side of
Eq.~(\ref{rate1}) is computed perturbatively to order $\alpha$ and in
principle to all orders in $\alpha_s$ by using real-time Feynman rules and
propagators. To {\em lowest order} in perturbation theory, the result for two
light quark flavors ($u$ and $d$) reads
\begin{eqnarray}
E\frac{dN(\tau)}{d^3p\,d^4x}\bigg|_{\rm LR}&=&\frac{2}{(2\pi)^3}
\int_{\tau_i}^{\tau}d\tau'' \int_{-\infty}^{+\infty}
\frac{d\omega}{\pi}\,\mathcal{R}(\omega)
\,\cos[(\omega-E)(\tau-\tau'')], \label{noneqratei}
\end{eqnarray}
where $\mathcal{R}(\omega)$ is the same as that given in
Eq.~(\ref{Ri}), but now with $n(q)=1/[e^{q/T_i}+1]$ being the
quark distribution function {\em at initial proper time} $\tau_i$,
at which the QGP reaches local thermal equilibrium.

In principle in the case of an expanding QGP under consideration, the photon
production rate given by Eq.~(\ref{noneqratei}) has to be supplemented by
kinetic equations that describe the evolution of the quark and gluon
distribution functions so as to setup a closed set of coupled equations.
However the assumption of the validity of (ideal) hydrodynamics entails that
the quarks and gluons form a perfectly coupled fluid. This in turn implies that
the mean free paths of the quarks and gluons are much shorter than the typical
wavelengths and the relaxation time scales are much shorter than the typical
time scales, i.e., the quark and gluon distribution functions adjust to local
thermal equilibrium instantaneously. Thus the assumption of the validity of
(ideal) hydrodynamics bypasses the necessity of the coupled kinetic equations:
quarks and gluons are in local thermal equilibrium at all times. Therefore,
within the framework of hydrodynamics we obtain the invariant photon production
rate by directly replacing the initial quark distribution function $n(q)$ in
Eq.~(\ref{Ri}) by the ``updated'' distribution function
$n[q,T(\tau'')]=1/[e^{q/T(\tau'')}+1]$ at proper time $\tau''>\tau_i$, where
$T(\tau'')$ is determined by the cooling law Eq.~(\ref{coolinglaw}). The
nonequilibrium invariant photon production rate that is consistent with the
underlying hydrodynamics is then given by
\begin{eqnarray}
E\frac{dN}{d^3p\,d^4x}\bigg|_{\rm
LR}[\tau,E,T(\tau)]&=&\frac{2}{(2\pi)^3}
\int_{\tau_i}^{\tau}d\tau''
\int_{-\infty}^{+\infty}\frac{d\omega}{\pi}\,
\mathcal{R}[\omega,T(\tau'')]\nonumber\\&&\times
\,\cos[(\omega-E)(\tau-\tau'')], \label{noneqrate}
\end{eqnarray}
with
\begin{eqnarray}
\mathcal{R}[\omega,T(\tau)]&=&\frac{20\,\pi^2\alpha}{3}
\int\frac{d^3q}
{(2\pi)^{3}}\Big\{2\big[1-({\hat{p}}\cdot{\hat{k}})
({\hat{p}}\cdot{\hat{q}})\big]n[q,T(\tau)]\bar{n}[k,T(\tau)]\nonumber\\
&&\times\,\delta(\omega+k-q)+\big[1+({\hat{p}}\cdot{\hat{k}})
({\hat{p}}\cdot{\hat{q}})\big]n[q,T(\tau)]
n[k,T(\tau)]\nonumber\\
&&\times\,\delta(\omega-k-q)\Big\},\label{R}
\end{eqnarray}
where $n[q,T(\tau)]=1/[e^{q/T(\tau)}+1]$ and $\bar{n}=1-n$. The momentum $q$
integrals in Eq.~(\ref{R}) are calculated in the LR frame and hence are
equivalent to those of the nonexpanding case above.

Before proceeding further, we emphasize two noteworthy features of
the non\-equilibrium invariant photon production rate: (i) The
photon production processes do {\em not} conserve energy. (ii) The
rate depends on (proper) time not only implicitly through the
local temperature but also explicitly. Furthermore, this explicit
(proper) time dependence is non-Markovian as clearly displayed in
Eq.~(\ref{noneqrate}). Whereas these two features seem to be
rather uncommon within the framework of the usual $S$-matrix
approach to transport in heavy ion collisions~\cite{greiner}, they
are {\em not} unusual in quantum kinetics of nonrelativistic
many-body systems~\cite{boyanmarkov,haug}. Indeed, energy
nonconserving transitions and memory effects which cannot be
explained by usual (energy conserving) Boltzmann kinetics have
been observed recently in the ultrafast spectroscopy of
semiconductors that are optically excited by a femtosecond laser
pulse~\cite{qk}.

The real-time kinetic approach when incorporated consistently with
hydrodynamics reveals clearly that photon production from an expanding QGP is
inherently a nonequilibrium quantum effect associated with the {\em finite
lifetime} of the QGP.

At RHIC and LHC energies the quark distribution function
$n[q,T(\tau)]$ depends on the proper time $\tau$ very weakly
through the temperature $T(\tau)$ within the lifetime of the QGP
phase, hence a Markovian approximation (MA) in which the
temperature in $\mathcal{R}[\omega,T(\tau'')]$ is taken at the
upper limit of the integral is reasonable and hence the memory
kernel may be simplified. In this Markovian approximation,
$\mathcal{R}[\omega,T(\tau'')]$ in Eq.~(\ref{noneqrate}) is
replaced by $\mathcal{R}[\omega,T(\tau)]$ and taken outside of the
$\tau''$-integral. Thus  Eq.~(\ref{noneqrate}) becomes
\begin{eqnarray}
E\frac{dN}{d^3p\,d^4x}\bigg|^{\rm MA}_{\rm
LR}[\tau,E,T(\tau)]&=&\frac{2}{(2\pi)^3}
\int_{-\infty}^{+\infty}d\omega\,\mathcal{R}[\omega,T(\tau)]\nonumber\\&&\times
\,\frac{\sin[(\omega-E)(\tau-\tau_i)]}{\pi(\omega-E)}.
\label{noneqrate2}
\end{eqnarray}
A computational advantage of this Markovian nonequilibrium production rate is
that it provides the ``updated'' quark distribution functions locally in time.
Physically, the motivation for this approximation is that the most important
aspect of the nonequilibrium effect is the nonconservation of energy (i.e.,
off-shellness) originated in the finite lifetime of the QGP, a feature that is
missed by the usual $S$-matrix calculation, while the proper time variation of
the temperature is a secondary effect and accounted for in the $S$-matrix
approach.

It is worth noting that a connection with the Boltzmann approximation can be
obtained by assuming completed collisions, i.e., taking the limit
$\tau_i\rightarrow -\infty$ in the argument of the sine function in
Eq.~(\ref{noneqrate2}) and using the approximation given by
Eq.~(\ref{goldenrule}). Consequently, the {\em lowest order} nonequilibrium
photon production rate vanishes in the Boltzmann approximation due to
kinematics. This highlights, once again,  that the usual approach to photon
production outlined in Sec.~\ref{sec:usualapproach} and used in the literature
corresponds  to the Boltzmann approximation and therefore fails to capture the
energy nonconserving and memory effects that occur during the transient stage
of evolution of the QGP.

We are now in a position to calculate the direct photon yield from a
longitudinally expanding QGP. In the CM frame the invariant production rate for
photons of four-momentum $P^{\mu}$ from a fluid cell with four-velocity $u^\mu$
can be obtained from Eq.~(\ref{noneqrate2}) through the replacement
$E\rightarrow P^{\mu}u_{\mu}$. In terms of the photon transverse momentum $p_T$
and rapidity $y$, one finds $P^{\mu}u_{\mu}=p_T\cosh(y-\eta)$. The photon yield
is obtained by integrating the nonequilibrium rate over the space-time
evolution of the expanding QGP. Assuming  a central collision of identical
nuclei, in the Markovian approximation we find the invariant nonequilibrium
photon yield to lowest order in perturbation theory to be given by
\begin{eqnarray}
\frac{dN}{d^2p_T\,dy}\bigg|^{\rm MA}_{\rm CM}&=&\pi R_A^2
\int_{\tau_i}^{\tau_f}d\tau\,\tau
\int_{-\eta_{\rm cen}}^{\eta_{\rm cen}}d\eta 
\,E\frac{dN}{d^3p\,d^4x}\bigg|^{\rm MA}_{\rm LR}
[\tau,P^{\mu}u_{\mu},T(\tau)],\label{yield}
\end{eqnarray}
where $R_A$ is the radius of the nuclei and $-\eta_{\rm cen}<\eta<\eta_{\rm
cen}$ denotes the central rapidity region within which Bjorken's hydrodynamical
model is valid.

As remarked above, at RHIC and LHC energies the quark distribution
function $n[q,T(\tau)]$ depends on the proper time $\tau$ very
weakly through the temperature $T(\tau)$ within the lifetime of
the QGP phase, therefore the resultant photon yield is expected to
qualitatively resemble the nonequilibrium photon yield from a
nonexpanding QGP studied in Ref.~\cite{boyanphoton}. This will be
numerically verified below. We note that the expanding and
nonexpanding cases differ mainly by the Jacobian $\tau$ in the
$\tau$ integral in Eq.~(\ref{yield}) that accounts for the
longitudinal expansion of the QGP, and by the replacement
$E\rightarrow P^{\mu}u_{\mu}$ in the argument of the invariant
photon production rate that accounts for the shift of the photon
energy under local Lorentz boost.

\begin{figure}[ht]
\begin{center}
\includegraphics[width=3.5in,keepaspectratio=true]{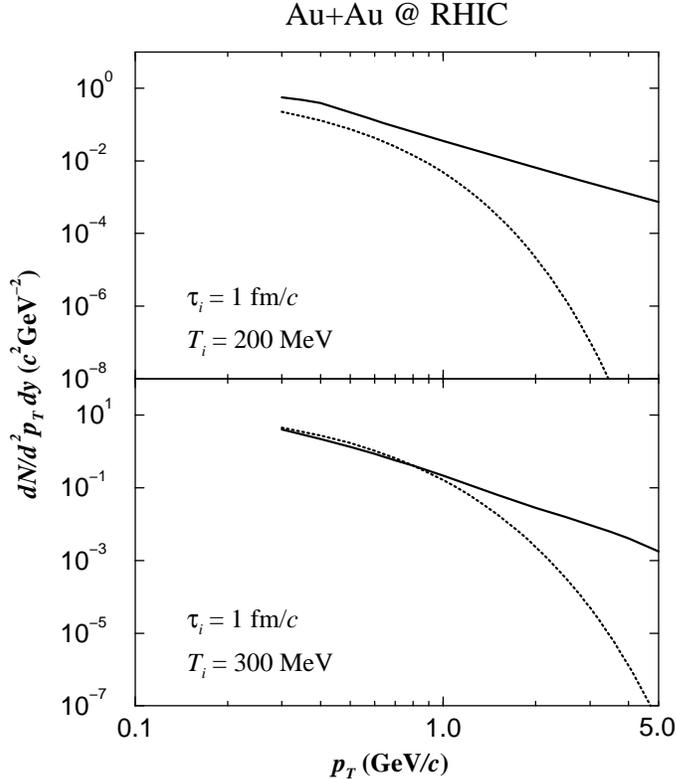}
\caption{Comparison of nonequilibrium (solid) and equilibrium
(dotted) photon yields at midrapidity ($y=0$) from a
longitudinally expanding QGP at RHIC energies with initial
conditions given by $\tau_i=1$ fm/$c$ and $T_i=200$ (top), 300
(bottom) MeV.} \label{fig:rhic}
\end{center}
\end{figure}

\subsection{Numerical analysis: central collisions at RHIC and LHC energies}

We now perform a numerical analysis of the nonequilibrium photon
yield and compare the results to the equilibrium one obtained from
higher order equilibrium rate calculations given by
Eq.~(\ref{eqrate}). The nonequilibrium photon yield in the
Markovian approximation given by Eq.~(\ref{yield}) contains a
four-dimensional integral that is performed numerically for the
values of parameters of relevance at RHIC and LHC energies.

For central ${}^{197}$Au$+{}^{197}$Au collisions at RHIC energies
$\sqrt{s}\sim 200A$ GeV, we take $R_A\simeq 1.2\,A^{1/3}$ fm
$\approx 7$ fm~\cite{lebellac} and $\eta_{\rm cen}=2$. The initial
thermalization time is taken to be $\tau_i=1$
fm/$c$~\cite{alam,bjorken}, the final proper time $\tau_f$ is
determined when the critical temperature for the quark-hadron
transition is reached at $T(\tau_f)\simeq 160$ MeV and is obtained
from the cooling law given by Eq.~(\ref{coolinglaw}) for a given
initial temperature $T_i$ at proper time $\tau_i$.

The nonequilibrium photon yield at midrapidity ($y=0$) in the
range of transverse momentum $T_i < p_T < 5$ GeV/$c$ is shown on a
log-log plot in Fig.~\ref{fig:rhic} for initial temperatures
$T_i=200$ and 300 MeV. For comparison we also plot the
corresponding equilibrium yield obtained by integrating
Eq.~(\ref{eqrate}) with the transformation to the CM frame as
specified by Eq.~(\ref{yieldSmtx}), and using the value of
$\alpha_s$~\cite{srivastava,alam,karsch}
\begin{equation}
\alpha_s[T(\tau)]=\frac{6\pi}{(33-2N_f)\ln[8\,T(\tau)/T_c]},
\end{equation}
with $N_f=2$. Furthermore, Figure~\ref{fig:rapidity} shows the
rapidity distribution of the nonequilibrium photon yield at
different values of $p_T$ for $T_i=200$ and $300$ MeV. Several
noteworthy features are gleaned from these figures:

\begin{figure}[t]
\begin{center}
\includegraphics[width=3.5in,keepaspectratio=true]{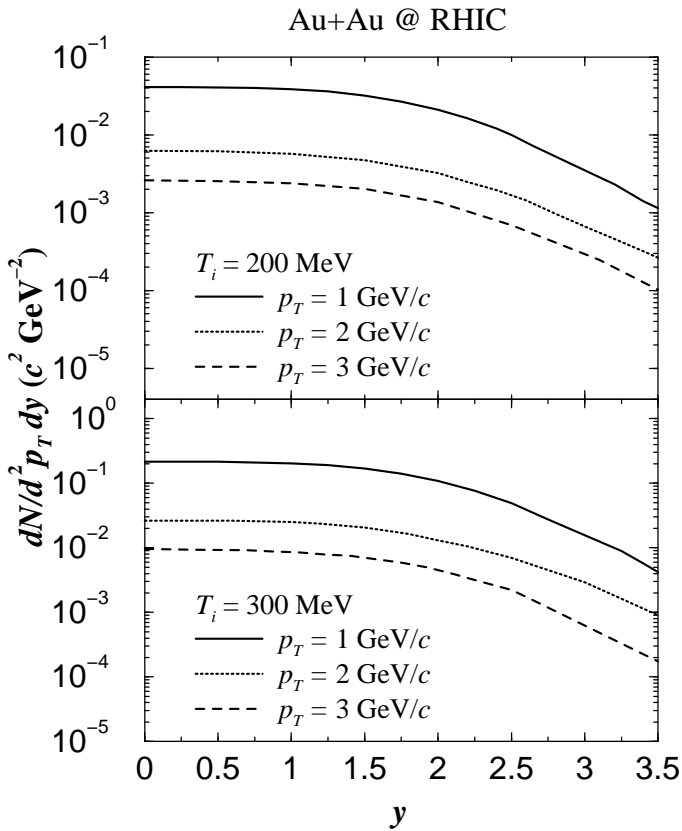}
\caption{Rapidity distribution of the nonequilibrium photon yield
at $p_T=1$ (solid), 2 (dotted), and 3 (dashed) GeV for a
longitudinal expanding QGP at RHIC energies with initial
conditions given by $\tau_i=1\,{\rm fm}/c$ and $T_i=200$ (top),
300 (bottom) MeV. The distribution is symmetric at $y=0$.}
\label{fig:rapidity}
\end{center}
\end{figure}

\begin{itemize}
\item[(i)]{Whereas the equilibrium yield dominates the total yield for small $p_T$,
the nonequilibrium yield becomes significantly dominant in the
range $p_T>1.0-1.5$ GeV/$c$. Perhaps coincidentally, this is the
range in which the CERN WA98 data for central Pb$+$Pb collisions
at SPS energies shows a distinct excess~\cite{WA98}.}

\item[(ii)]{While the equilibrium yield leads to a transverse momentum distribution
that falls off approximately with the Boltzmann factor
$e^{-p_T/T_i}$, the nonequilibrium yield is {\em not} Boltzmann
suppressed {\em and} falls off algebraically. It is found
numerically that for $p_T\gg T_i$ and within the region $1
\lesssim p_T \lesssim 5$ GeV/$c$ the nonequilibrium yield at
midrapidity falls off with a power law $p_T^{-\nu}$ with
$\nu\simeq 2.47$ and $2.77$ for $T_i=200$ and 300 MeV,
respectively. This is a remarkable consequence of the fact that
bremsstrahlung is the dominant process. As discussed in detail in
Ref.~\cite{boyanphoton} at large energies the dominant process is
bremsstrahlung corresponding to the contribution from the term
$n(q)[1-n(k)]$ with $k = |{\vec p}+{\vec q}|$ in Eq.~(\ref{Ri}).
For large photon energy $p$ the important contribution, which is
not exponentially suppressed  arises from the small $q$ region. }

\item[(iii)]{A central plateau in the range of rapidity $y\lesssim 2$ is seen
clearly. The rapidity distribution begins to bend down when $y\simeq\eta_{\rm
cen}$, i.e., when the photon rapidity probes the fragmentation region.}

\item[(iv)] {The numerical analysis of the expanding case reveals
features very similar to those found in the nonexpanding case
studied in Ref.~\cite{boyanphoton}, where we estimate that for
$p_T \geq 1.5$ GeV/$c$ the higher order equilibrium contribution
to the direct photon yield (which grows linearly in time) becomes
of the same order as the lowest order nonequilibrium contribution
(which grows at most logarithmically in time) only if the lifetime
of the QGP is of the order longer than 1000 fm/$c$. Therefore, we
conclude that these nonequilibrium effects dominate during the
lifetime of the QGP in a realistic ultrarelativistic heavy ion
collision experiment.}
\end{itemize}

\begin{figure}[t]
\begin{center}
\includegraphics[width=3.5in,keepaspectratio=true]{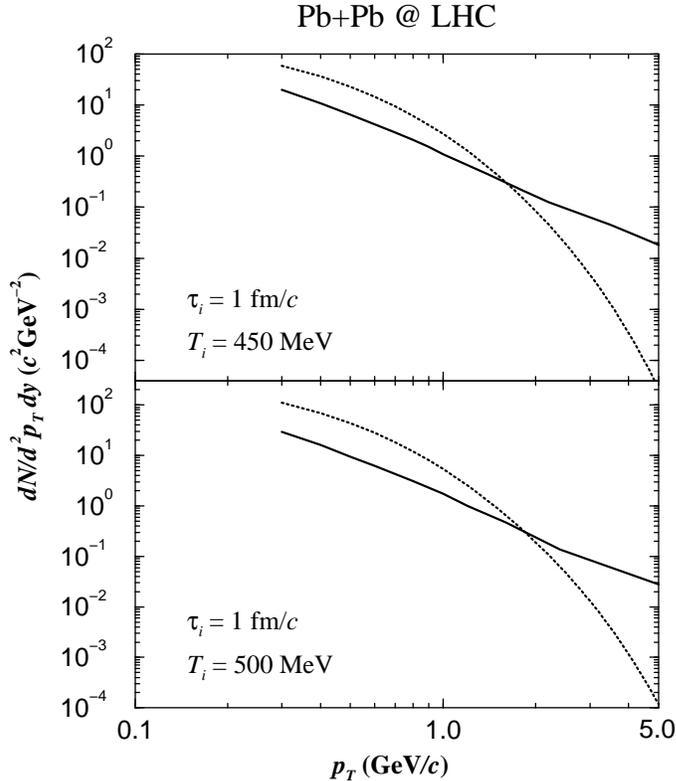}
\caption{Comparison of nonequilibrium (solid) and equilibrium
(dotted) photon yields at midrapidity ($y=0$) from a
longitudinally expanding QGP at LHC energies with initial
conditions given by $\tau_i=1$ fm/$c$ and $T_i=450$ (top), 500
(bottom) MeV.} \label{fig:lhc}
\end{center}
\end{figure}

Similar numerical analysis can be performed for central
${}^{208}$Pb$+{}^{208}$Pb collisions at higher LHC energies
$\sqrt{s}\sim 5500A$ GeV, for which we take $R_A\approx 7$ fm,
$\eta_{\rm cen}=5$ and $\tau_i=1$ fm/$c$~\cite{alam,bjorken}. The
comparison of nonequilibrium and equilibrium photon yields at
midrapidity is displayed in Fig.~\ref{fig:lhc} for $T_i=450$ and
500 MeV. The dominance of the nonequilibrium yield at high $p_T$
remains but now at higher transverse momentum $p_T\gtrsim 2$ GeV.
This can be understood as a consequence of the longer QGP lifetime
resulting from higher initial temperature at LHC energies.
Furthermore, it is found numerically that for $1 \lesssim p_T
\lesssim 5$ GeV/$c$ the nonequilibrium yield at midrapidity falls
off with a power law $p_T^{-\nu}$ with $\nu\simeq 2.52$ and $2.56$
for $T_i=450$ and 500 MeV, respectively.

These results indicate a clear manifestation of the nonequilibrium
aspects of direct photon production associated with a transient
QGP of finite lifetime. The most experimentally accessible signal
of the nonequilibrium effects revealed by this analysis is the
{\em power law} falloff of the transverse momentum distribution
for direct photons in the range $1 \lesssim p_T \lesssim$ 5
GeV/$c$ with an exponent $2.5 \lesssim \nu \lesssim 3$ for
temperatures expected at RHIC and LHC energies. Our numerical
studies reveal that this exponent increases with initial
temperature and therefore with the initial energy density of the
QGP and the total multiplicity rapidity distribution
$dN_\pi/dy$~\cite{bjorken}. This could be a clean experimental
{\em nonequilibrium signature} of a transient QGP since the photon
distribution from the hadronic gas is expected to feature a
Boltzmann type exponential suppression for $p_T \gg
T_i$~\cite{kapusta,sollfrank,srivastava,alam}.

\section{Conclusions and discussions}

Our goal in this article is to search for clear experimental
signatures of direct photons associated with nonequilibrium
aspects of the transient quark-gluon plasma created in
ultrarelativistic heavy ion collisions at RHIC and LHC energies.

We argue that the usual $S$-matrix approach to direct photon production from an
expanding nonequilibrium QGP has conceptual limitations. Instead, we introduce
a real-time kinetic approach that allows a consistent treatment of photon
production from a transient nonequilibrium state of {\em finite lifetime}.

We focus on obtaining the direct photon yield from a thermalized
QGP undergoing Bjorken's hydrodynamical expansion in central
Au$+$Au collisions at RHIC ($\sqrt{s}\sim 200A$ GeV) and LHC
($\sqrt{s}\sim 5500A$ GeV) energies. The lifetime of a QGP for
these collisions is of order $\lesssim 10-30$ fm/$c$. We find that
(anti)quark bremsstrahlung $q(\bar{q})\rightarrow
q(\bar{q})\gamma$ and quark-antiquark annihilation
$q\bar{q}\rightarrow\gamma$, {\em both of lowest order in
$\alpha$}, dominate during such short time scales, with
bremsstrahlung dominating for $p_T \gg T_i$. The contribution from
these processes is a consequence of the transient nature and the
finite lifetime of the QGP. As compared to the equilibrium rate
calculations, these  processes lead to a substantial enhancement
in direct photon production for $1-2 \lesssim p_T \lesssim$ 5
GeV/$c$ near midrapidity. In striking contrast with the
equilibrium calculation that predicts an exponential suppression
of the transverse momentum distribution for $p_T \gg T_i$ ($T_i$
is the initial temperature), the nonequilibrium processes lead to
a {\em power law} behavior instead. We find that at RHIC and LHC
energies the direct photon transverse momentum distribution near
midrapidity is of the form $p^{-\nu}_T$ with $2.5 \lesssim \nu
\lesssim 3$ for $1 \lesssim p_T \lesssim$ 5 GeV/$c$ and that
photon rapidity distribution (for fixed $p_T$) is almost flat in
the interval $|y| \lesssim \eta_{\rm cen}$, where
$|\eta|<\eta_{\rm cen}$ denotes the central rapidity region of the
QGP. The exponent $\nu$ is numerically found to increase with the
initial temperature, hence increases with the total multiplicity
rapidity distribution $dN_\pi/dy$, which is an experimental
observable.

Thus, as the main conclusion, we propose that direct photons could
potentially provide distinct experimental signatures of the
transient nonequilibrium QGP created at RHIC and LHC energies,
both in the form of a large enhancement at $1-2 \lesssim p_T
\lesssim$ 5 GeV/$c$ as well as a power law transverse momentum
distribution $p^{-\nu}_T$ with an exponent $\nu$ that is within
the range $2.5 \lesssim \nu \lesssim 3$ and increases with total
multiplicity rapidity distribution $dN_\pi/dy$.


\subsection*{Discussions.}
\begin{itemize}
\item[(i)]{Our study of direct photon production focuses solely on the QGP phase
and neglects contributions from pre-equilibrium stage as well as the
mixed and hadronic phases, because we focus on a comparison between the
usual equilibrium approach and the kinetic real-time approach. While
the usual equilibrium approach treats the thermalized state as
stationary, the real time kinetic approach  allows to follow the time
evolution of the density matrix as an initial value problem consistent
with hydrodynamics. Furthermore, as discussed elsewhere in the
literature, since most of the high-$p_T$ photons originate from the
very early, hot stage in the QGP
phase~\cite{mclerran,kapusta,ruuskanen,gale} and photons produced in
the mixed phase as well as the subsequent hadronic phase are mainly in
the lower-$p_T$ region~\cite{alam}, the nonequilibrium yield from the
QGP phase contributes dominantly to the total high-$p_T$ photons.}

\item[(ii)]{While we have provided a systematic real-time description compatible
with the initial value problem associated with hydrodynamic evolution,
more needs to be understood for a complete description of {\em all} the
different stages. The parton cascade approach to describe the early
pre-equilibrium stage after the collision is an important first step in
a full microscopic description, but perhaps a more consistent
description of the initial condition for QGP formation must be based on
the recent notions of a color glass condensate~\cite{glass,raju}.
Hence, a complete treatment of the direct photon yield must, in
principle, begin from the initial stage, possibly a color glass
condensate, and obtain the real-time evolution of photon production.
Clearly there must be many more advances in this field before such a
program becomes feasible. The finite-temperature equilibrium
calculations {\em assume} that the equilibrium thermal state always
prevailed, thus ignores completely not only the initial stages but also
the time evolution. Our approach while incorporating the time evolution
during the stage of local thermodynamic equilibrium consistently, also
neglects the initial stage. However, the advantage of our method is
that if there is an estimate for the photon distribution at the onset
of thermalization we could evolve the full quantum kinetic equation,
including the higher order corrections in $\alpha$ as discussed in the
text. We have focused on direct photons from a transient QGP for a
direct comparison with equilibrium calculations, but obviously photons
will continue to be produced during the mixed and hadronic phases. A
detailed understanding of the phase transition as well as the hadronic
photon production matrix elements is necessary for a more reliable
estimate of the potential nonequilibrium effects {\em after}
hadronization.}

\item[(iii)]{Although we do not claim to have provided a completely detailed
understanding of potential nonequilibrium effects due to the lack
of knowledge of initial conditions in heavy ion collisions, we do
claim that our approach provides a more systematic description of
the finite-lifetime effects associated with a transient QGP. These
effects lead to a distinct experimental prediction: a power law
falloff of the distribution $dN/d^2p_T dy$ near the central
rapidity region which is distinguishable from a thermal tail for
$1-2 \lesssim p_T \lesssim 5$ GeV/$c$. We also note that the
parton cascade results of Srivastava and Geiger~\cite{srivageiger}
also seem to reveal a power law falloff of the photon distribution
in this range of momentum with a similar amplitude and comparable
exponent, as can be gleaned from Figs.~2 and 3 of
Ref.~\cite{srivageiger}. Perhaps these two effects, namely the
pre-equilibrium prompt photons and the direct photons from a
hydrodynamically expanding QGP with a finite lifetime cannot be
distinguished experimentally. Nevertheless, we emphasize that a
power law departure from a thermal tail in the direct photon
spectrum may very well be explained by nonequilibrium effects,
either by a finite QGP lifetime as advocated in this article or by
prompt photons from the pre-equilibrium stage}.

\item[(iv)]{As many recent investigations~\cite{chemnoneq1,chemnoneq2} have
suggested that the QGP produced at RHIC and LHC energies is not
expected to be in local chemical equilibrium, i.e., the
distribution functions of quarks and gluons will probably be
undersaturated. An important extension of our work  will be  the
study of nonequilibrium effects on direct photon production from a
chemically nonequilibrated QGP.}

\item[(v)]{{\bf Transient QGP vs high energy
particle collisions}. An important and very relevant question is
that why not treat high energy particle collisions in the same
manner, i.e., with the real-time evolution rather than with the
$S$-matrix approach. The answer to this question hinges on the
issue of time scales. In a typical high energy particle collider
experiment, the colliding ``beams'' are actually bunches or
packets with a typical spatial extent of order $1-10$ cm and hence
the typical total time interval for the collision is of order
$10^{-9}$ sec~\cite{exp}. This time scale is many orders of
magnitude longer than the typical hadronic interaction time scale
$\sim 10^{-23}$ sec, thus taking the infinite time limit is amply
justified. This, of course, is the basis for using the $S$-matrix
calculation in terms of asymptotic {\sl in} and {\sl out} states
(at $t=\mp\infty$): the total interaction time ($t_f-t_i$) is
much, much longer than the typical hadronic time scale. With
regard to the heavy ion collision, the consensus is that after an
initial pre-equilibrium stage, a locally thermalized QGP is
formed. It then evolves hydrodynamically, hadronizes and
eventually the freeze-out of the hadronic gas ensues. Current
theoretical estimates at RHIC energies indicate a total time
between formation and freeze-out of order $50-100$ fm/$c$, with a
QGP phase lasting for about $10$ fm/$c$. These time scales, at
least the lifetime of the QGP,  are {\em not} several orders of
magnitude larger than the hadronic interaction time scale, thus
the infinite time limit taken in the $S$-matrix calculation is at
best questionable. Therefore while in typical particle collider
experiments the $S$-matrix approach is valid (as has been
demonstrated by over half a century of experiments!), the
transient QGP with a finite lifetime of the order of the hadronic
time scales merits a {\em different} treatment, which is the point
of this article.}

\item[(vi)]{An important and related aspect that requires further investigation
is the finite size of the QGP. Much in the same way as the finite lifetime
introduces nonequilibrium effects associated with energy nonconserving
transitions, we expect that the finite size $\sim 7$ fm of the QGP will
introduce uncertainty in the momentum of the emitted particles. This is clearly
an important topic that deserves further and deeper study but is beyond the
scope of this article. However, at this stage we speculate that if such effects
are present they could bear an imprint in transverse flow.}
\end{itemize}
Work along many of these directions is currently in progress.

\section*{Acknowledgements}
We thank H.\ J.\ de Vega, J.\ I.\ Kapusta, B.\ M\"uller, D.\ H.\
Rischke, and H.\ A.\ Weldon  for useful discussions. The work of
D.B.\ and S.-Y.W.\ was supported in part by the US NSF under
grants PHY-9605186, PHY-9988720, and INT-9905954. S.-Y.W.\ would
like to thank the Andrew Mellon Foundation for partial support.
The work of K.-W.N.\ was supported in part by the ROC NSC under
grants NSC89-2112-M-259-008-Y and NSC89-2112-M-001-001.

\appendix
\section{Photon production rate and current correlation
function}

The equilibrium thermal photon production rate is related to the Fourier
transform of the thermal expectation value of the quark electromagnetic current
correlation function~\cite{feinberg,mclerran,gale}. In this appendix we show
that the photon production rate given by Eq.~(\ref{a:rate1}) is related to the
Fourier transform of the {\em nonequilibrium expectation value} of the quark
electromagnetic current correlation function.

In terms of the (Abelian) gauge invariant fields the interaction Lagrangian
between quarks and photons reads~\cite{boyanQED}
\begin{equation}
\mathcal{L}_{\rm int}=e\,{\vec J}\cdot{\vec A}_T,
\end{equation}
with ${\vec A}_T$ the transverse component of the photon field and ${\vec J}$
the spatial component of the quark electromagnetic current
\begin{equation}
{\vec J}=\sum^{N_f}_{f=1}\,e_f\,\bar{\psi}_f{\vec\gamma}\psi_f.
\end{equation}
The invariant photon production rate given by Eq.~(\ref{a:rate1}) can be
written in terms of ${\vec J}$ as
\begin{eqnarray}
E\frac{dN(t)}{d^3p\,d^4x}&=&\lim_{t'\rightarrow t} \frac{e}{2(2\pi)^3}
\left(\frac{\partial}{\partial \tau'}-iE\right) 
\left\langle{\vec J}^-(-{\vec p},t)\cdot{\vec A}^+_T({\vec p},t')
\right\rangle+{\rm c.c.}.\label{app:rate1}
\end{eqnarray}
It is convenient to introduce the Fourier transform of the transverse component
of the quark current correlation function $W_T^{\mbox{\scriptsize
\raisebox{1.8pt}{\raisebox{1.8pt}{$\scriptscriptstyle>$}
\raisebox{-1pt}{$\scriptscriptstyle\!\!\!\!\!\!<$}}}}(\omega,p)$ defined by
\begin{eqnarray}
\frac{1}{2}\,\mathcal{P}^{ij}({\hat p})\left\langle J^{i+}({\vec
p},t)J^{j-}(-{\vec p},t')\right\rangle &=&
\int^{+\infty}_{-\infty}
\frac{d\omega}{2\pi}\,W_T^<(\omega,p)\,e^{-i\omega(t-t')},\\
\frac{1}{2}\,\mathcal{P}^{ij}({\hat p})\left\langle J^{i-}({\vec
p},t)J^{j-}(-{\vec p},t')\right\rangle &=&
\int^{+\infty}_{-\infty}
\frac{d\omega}{2\pi}\,\big[W_T^>(\omega,p)
\theta(t'-t)\nonumber\\
&&+\,W_T^<(\omega,p)\theta(t-t')\big]\, e^{-i\omega(t-t')},
\end{eqnarray}
where $\mathcal{P}^{ij}({\hat p})=\delta^{ij}-\hat{p}^i\hat{p}^j$
is the transverse projector. A diagrammatic expansion [see
Fig.~\ref{fig:loop}(a)] shows that to order $\alpha$ and to all
orders in $\alpha_s$ Eq.~(\ref{app:rate1}) can be expressed in
terms of $W^<(\omega,p)$ as
\begin{eqnarray}
E\frac{dN(t)}{d^3p\,d^4x}&=&\frac{e^2}{(2\pi)^3}
\int_{t_i}^{t}dt''\int^{+\infty}_{-\infty}
\frac{d\omega}{2\pi}\,W_T^<(\omega,p) 
\,e^{-i(E-\omega)(t-t'')}+{\rm c.c.},\label{app:rate2}
\end{eqnarray}
where use has been made of the vacuum photon propagators. For a
QGP with an initial state at time $t_i$ described by a {\em
thermal} density matrix, one can show that the perturbative
Kubo-Martin-Schwinger (KMS) condition holds to all orders in
$\alpha_s$, i.e.,
\begin{equation}
W_T^<(\omega,p)=e^{-\omega/T_i}\,W_T^>(\omega,p),
\end{equation}
where $T_i$ is the initial temperature. Consequently, one finds to all orders
in $\alpha_s$ that $W_T^<(\omega,p)$ is a real quantity and related to the
imaginary part of the retarded photon self-energy. Therefore
Eq.~(\ref{app:rate2}) becomes
\begin{eqnarray}
E\frac{dN(t)}{d^3p\,d^4x}&=& \frac{2e^2}{(2\pi)^3}
\int_{t_i}^{t}dt''\int^{+\infty}_{-\infty}
\frac{d\omega}{2\pi}\,W_T^<(\omega,p) 
\,\cos[(\omega-E)(t-t'')],\label{app:rate3}
\end{eqnarray}
which is correct to order $\alpha$ and to all orders in $\alpha_s$. This result
generalizes that of Refs.~\cite{feinberg,mclerran,gale} to the nonequilibrium
situation in the real-time formulation. Upon comparing Eq.~(\ref{app:rate3})
with Eq.~(\ref{a:noneqratei}), one finds
\begin{equation}
\mathcal{R}(\omega)=\frac{e^2}{2}\left.W_T^<(\omega,p)\right|_{\text{one-loop}}.
\end{equation}


\end{document}